# STATUS REPORT ON THE 5 MeV IPHI RFQ

R. Ferdinand, P-Y. Beauvais, R. Duperrier, A. France, J. Gaiffier,
J-M. Lagniel, M. Painchault, F. Simoens, CEA-Saclay, DSM-DAPNIA-SEA
P. Balleyguier, CEA-Bruyères le Châtel, DAM

*Abstract*

A 5-MeV RFQ designed for a proton current up to 100-mA CW is now under construction as part of the High Intensity Proton Injector project (IPHI). Its computed transmission is greater than 99 %. The main goals of the project are to verify the accuracy of the design codes, to gain the know-how on fabrication, tuning procedures and operations, to measure the output beam characteristics in order to optimise the higher energy part of the linac, and to reach a high availability with minimum beam trips. A cold model has been built to develop the tuning procedure. The present status of the IPHI RFQ is presented.

## 1 INTRODUCTION

Over the last 10 years, in-depth studies have been carried out on the feasibility of high-power proton accelerators capable of producing beams of several tens of MW. With heavy targets, such beams can produce extremely intense spallation neutron flux. Several applications could benefit from the performance of this new generation of high-power proton accelerators [1]: spallation neutron sources for condensed matter studies, hybrid reactors for nuclear waste transmutation, neutrino and muon factories, technological irradiation tools, production of radioactive ion beams, production of radioisotopes, etc.

IPHI ("Injector of Protons for High-Intensity beams") is a 1 MW low energy prototype, which could be used as front end for such high-power proton accelerators [2]. This demonstrator is made up of the SILHI ECR source able to deliver more than 100 mA CW at 95 keV, the 5 MeV RFQ and a 10 MeV DTL. IPHI is designed to operate up to 100% duty factor (CW).

## 2 RFQ BEAM DYNAMICS

The input energy of 95 keV results of a compromise between RFQ length, source reliability and space-charge control. The 5 MeV output energy results of a compromise between cavity length, feasibility of the DTL using EM quadrupoles, and high beam transmission. The use of existing klystron at 352.2 MHz leads to an optimum size of the cavity. The design current of 100 mA has been selected to reach a high reliability at the lower currents needed by the different applications. The expected normalised *rms* emittance from the source is 0.2 $\pi$.mm.mrad. Nevertheless, a safety margin is taken using 0.25 $\pi$.mm.mrad in beam dynamics calculations.

The maximum electric field has been limited to 1.7 Kp (31.34 MV/m) taking into account experiences with the CRITS Experiment at Los Alamos and RFQs operated at Saclay in the past. The RFQ cavity length was set to 8 m. Great care was taken on lost particles in the RFQ cavity. The final design has been selected to avoid localised and high-energy losses (activation), and to provide the highest transmission avoiding any bottle neck [3]. Many beam dynamics computations including error studies have been done using several complementary codes (PARMTEQM, TOUTATIS [4,6], and LIDOS.RFQ [5]). Table 1 gives the main parameters of the RFQ.

Table 1 : IPHI RFQ parameters

| Structure | 4 vanes |
|---|---|
| Frequency | 352.2 MHz |
| Total length | 8 m (8 sections) |
| Resonant coupling sections | 4 |
| Input/output Energy | 95 keV / 5 MeV |
| Input beam characteristics | 100 mA/0.25 $\pi$.mm.mrad |
| Mean aperture ($R_0$) | 3.7 - 5.3 mm |
| Modulation (m) | 1 - 1.75 |
| Vanes voltage | 87 - 123 kV (1.7 Kp) |
| PARMTEQM transmission | 99.2 % (accel. particles) |
| Beam Power | 490 kW |
| Total expected power | 1650 kW |
| Stored energy | 5.3 J |

The new TOUTATIS code [6] allows to take into account the field errors in the coupling section as well as the mechanical defaults (vane extremities displacement...). The LIDOS.RFQ code allowed to establish the required machining precision. Lots of errors were simulated with the coupling gaps to ensure a "faults tolerant" design.

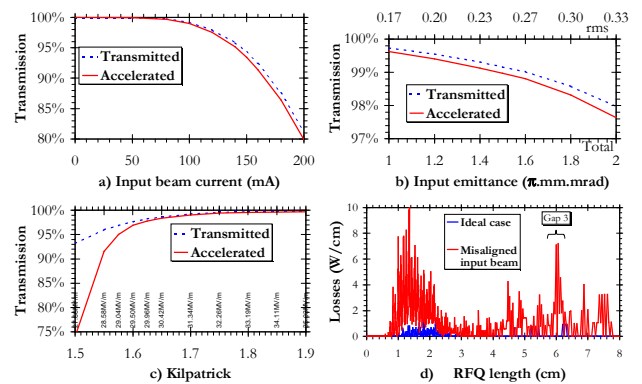

Figure 1: Transmission versus input beam current (a), input beam emittance (b), and rf field (c). Power deposition due to beam losses along RFQ (d)

## 3 RFQ CAVITY

The structure is made up of 8 one-meter long sections accurately machined and brazed then assembled with a resonant coupling every 2 meters (similar to the LEDA design [8]). Details of the "Main steps for fabrication of the IPHI RFQ" are published in these proceedings (THD03, M Painchault *et al.*).

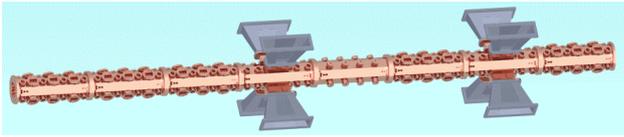

Figure 2: IPHI RFQ Artistic view with the 8 RF ports

The coupling plates allow a damping of the rf longitudinal parasitic modes and the introduction of fingers to push away the dipolar modes. The RF design of the cavity requires intensive 3D simulations and developments [7] with cross-checking on the cold model. The field will be tuned using 128 tuners equally distributed along the RFQ. The 8 Thomson RF windows are already on hand, similar windows have been successfully tested up to 700 kW at LANL for LEDA [8]. The low level RF is still under definition. A pick-up will be used for the fast phase and amplitude control using DSP. The slow frequency tuning will be done using the cooling system of the cavity based on the LEDA design [8]. The inlet water temperature is 10°C/50°F with water flows tuneable up to 6 m/s. An erosion/corrosion analysis is presently done.

Great care has been put into the optimization of the pumping system. Two of the 1-m long sections are dedicated to the rf feed while all the remaining sections are equipped with a total of 72 pumping ports carefully designed to maximize the pumping speed. The running pressure is expected to be $8 \cdot 10^{-6}$ Pa.

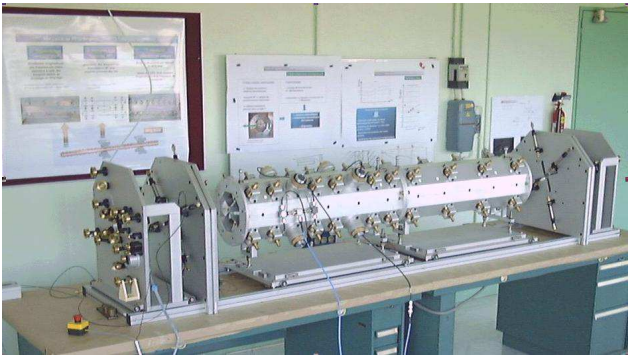

Figure 3: View of the RFQ cold model.

## 4 COLD MODEL

### 3.1 Objectives

The fabrication of a RFQ cold model started at the beginning of 1999. The aluminium cavities have been designed close to those of the final RFQ with possible fast adjustments of the geometry to test different end vane configurations, consequences of vane position errors... It is now mainly used to develop the RF tuning procedures and the associated hardware and software. Some of the development will be used to help on vane positioning before the brazing step with an expected precision better than $10^{-5}$ m.

### 3.2 Design

The 1:1 scale cold model has been designed using SUPERFISH for the main region and MAFIA for the end regions. The transverse section is composed of flat faces only, the tips of the electrodes are circular ($\rho/r_0 = 0,85$). The design has been done for a resonance frequency of 350.7 MHz with all the tuners flush mounted and 352.2 MHz with the tuners 5 mm inside the cavity.

### 3.3 Modularity

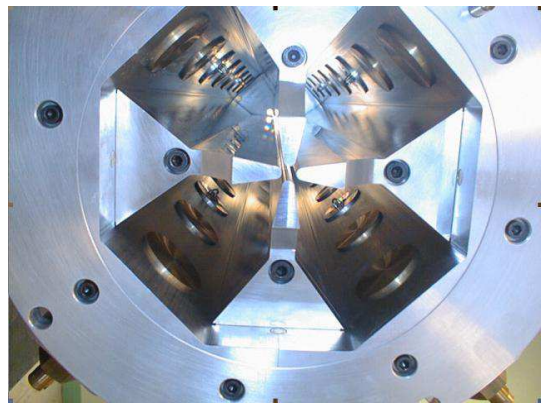

Figure 4 : transverse view of the RFQ cold model.

The cold model consists of octagonal 1-meter long sections in which the 4 electrodes are screwed. Three different kinds of end pieces may be screwed to the main electrodes: input/output beam, coupling region and plain pieces. This last type allows to form a 2 meters long RFQ with a continuous electrode. Three different kinds of segments can be used :

1- "pumping" type with the same slug tuners distribution as the final pumping segments,
2- "RF coupling" type with one rectangular hole per quadrant to allow the study of RF power coupling through irises,
3- "Tuners over-equipped" type with 8 tuners per quadrant instead of 4 for an accurate study of the voltage law tuning.

One "pumping" section and one "Tuners over-equipped" section are tested since August 99. The measured resonance frequencies with the tuners at the nominal position are respectively 350.054 MHz and 351.24 MHz, the relative error is less than $3.10^{-3}$. The lower 2 dipole modes are about 700 kHz apart. Four more segments will be tested soon.

An elaborated pulley system guides the bead on different path through the 4 quadrants allowing a comparison of the magnetic and electric fields measurement in several locations (see Figure 5).

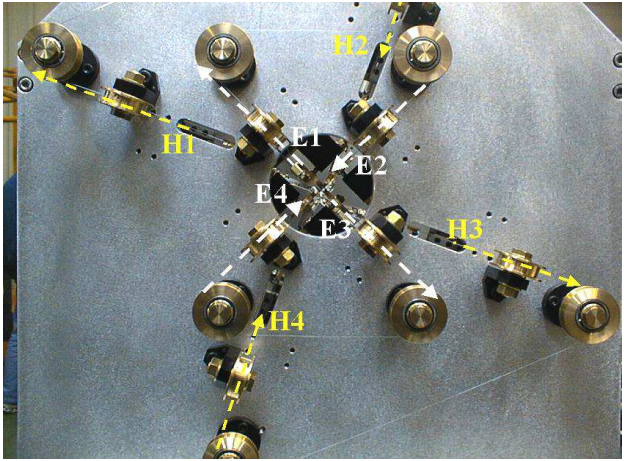

Figure 5: pulley system of the cold model

### 3.4 RF diagnostics

The fields are measured using the common bead–pull perturbation method. A DC motor drives the bead and a vector network analyser (VNA) measures the phase of the transmission coefficient (s21). Operation of the DC motor and the VNA is fully automatic, both being driven by a LabView program on PC. Measurements are readily available in data files.

### 3.5 Measurement analysis

The data are then treated through a Matlab code. The first step is the conversion of the measured phases into voltages versus position for each quadrant, all expressed in arbitrary units since no attempt is made to derive the polarisability of the bead or the phase versus frequency slope of the s21. Smoothing and windowing produce direct usable data. The second step is the analysis of the data from the 4-quadrants. The RFQ is modelled as a 4-wire line system. The spectral theory of differential operators is used to relate the measured voltages along the line to the physical parameters describing the whole RF circuit (the parallel capacitances $C_i$ or inductances $L_i$ of each quadrant $i$ versus the position and the end loads).

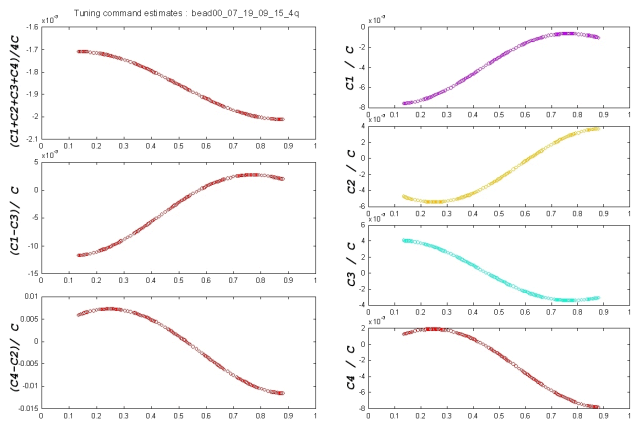

Figure 6: Estimated capacitances
(order = 2 for Q, S and T eigen modes)

Figure 6 shows the result of a 4-quadrant analysis. The magnetic field of the "over-equipped" type segment has been perturbed with a titanium bead guided close to flush mounted tuners. The left hand side figures are plots of modal combinations of capacitances over the theoretical capacitance C versus the longitudinal position z [m]. The plots at right are the $C_i/C$ for each quadrant. All the curves are well within plus and minus 1% indicating a very accurate machining and positioning of the 4 quadrants. The measurement of the capacitances is reproducible within $2.10^{-3}$.

## 5 CONCLUSION

The detailed design of the IPHI RFQ is now nearly completed. Great progress has been done on the RF tuning procedures using the cold model. The construction of a one-meter long copper prototype will be finished in September. The delivery of the first RFQ section is expected for the end of 2000 and the 8 sections must be available mid 2002. The assembly will start before the reception of the last section as shown in the planning below. The first beam is expected late 2002 / beginning 2003.

|  | 2000 | 2001 | 2002 | 2003 |
|---|---|---|---|---|
|  | 1 2 3 4 | 1 2 3 4 | 1 2 3 4 | 1 2 3 4 |
| IPHI site availability |  |  |  |  |
| Cooling system overhaul |  |  |  |  |
| Power supply distribution overhaul |  |  |  |  |
| Source/LEBT settling in |  |  |  |  |
| RFQ/rf/HEBT/BS assembly |  |  |  |  |
| RFQ conditioning |  |  |  |  |
| 5 MeV pulsed operation |  |  |  |  |
| 5 MeV CW operation |  |  |  |  |